\documentclass[a4paper,fleqn,usenatbib]{mnras}
\usepackage{newtxtext,newtxmath}
\usepackage[T1]{fontenc}
\usepackage{ae,aecompl}
\usepackage{graphicx}	
\usepackage{amsmath}	
\usepackage{amssymb}	
\usepackage{booktabs}	
\usepackage{url}

\newcommand{\specline}[3]{{#1}\,{\sc #2}\:{#3}}

\usepackage{tablefootnote,footnotehyper} 
\usepackage{array}
\newcolumntype{H}{>{\setbox0=\hbox\bgroup}c<{\egroup}@{}} 

\newcommand{\paperi}{\href{https://ui.adsabs.harvard.edu/abs/2017AJ....154...31G/abstract}{Paper\,I}}

\title[$\lambda$\,Boo stars, Southern Survey II]{The Discovery of lambda Bootis Stars -- The Southern Survey II}

\author[Simon J. Murphy et al.]{
Simon J. Murphy$^{1,2}$\thanks{E-mail: simon.murphy@sydney.edu.au (SJM)}, Richard O. Gray$^{3}$, Christopher J. Corbally$^{4}$, \and Charles Kuehn$^{1,5}$
Timothy R. Bedding$^{1,2}$ and Josiah Killam$^{3}$
\\
$^{1}$ Sydney Institute for Astronomy (SIfA), School of Physics, University of Sydney, NSW 2006, Australia\\
$^{2}$ Stellar Astrophysics Centre, Department of Physics and Astronomy, Aarhus University, 8000 Aarhus C, Denmark\\
$^{3}$ Department of Physics and Astronomy, Appalachian State University, Boone, NC 26808, USA\\
$^{4}$ Vatican Observatory Research Group, Steward Observatory, Tucson, AZ, 85721-0065 USA\\
$^{5}$ Department of Physics and Astronomy, University of Northern Colorado, Greeley, CO 80639, USA\\
} 

\date{Accepted XXX. Received YYY; in original form ZZZ}
\pubyear{2019}
\begin{document}

\label{firstpage}
\pagerange{\pageref{firstpage}--\pageref{lastpage}}
\maketitle

\begin{abstract}
The $\lambda$\,Boo stars are chemically peculiar A-type stars whose abundance anomalies are associated with the accretion of metal-poor material. We searched for $\lambda$\,Boo stars in the southern hemisphere in a targeted spectroscopic survey of metal-weak and emission-line stars. Obtaining spectra for 308 stars and classifying them on the MK system, we found or co-discovered 24 new $\lambda$\,Boo stars. We also revised the classifications of 11 known $\lambda$\,Boo stars, one of which turned out to be a chemically normal rapid rotator. We show that stars previously classified in the literature as blue horizontal branch stars or emission-line A stars have a high probability of being $\lambda$\,Boo stars, although this conclusion is based on small-number statistics. Using WISE infrared fluxes, we searched our targets for infrared excesses that might be attributable to protoplanetary or debris discs as the source of the accreted material. Of the 34 $\lambda$\,Boo stars in our sample, 21 at various main-sequence ages have infrared excesses, confirming that not all $\lambda$\,Boo stars are young.
\end{abstract}




\section{Introduction}
\label{sec:intro}

Long-standing puzzles in astrophysics often contain clues on physics that is missing from stellar models. The $\lambda$\,Boo stars are one such puzzle. They are chemically peculiar A or F-types stars first identified as a distinct class in the 1950s \citep{slettebak1952,slettebak1954}, and a complete explanation for their peculiarity is still lacking despite recent efforts \citep{jura2015,kamaetal2015,jermyn&kama2018}. They are characterised by metal weaknesses with a specific chemical abundance profile. Refractory elements such as magnesium and iron-peak elements are underabundant by $-$0.5 to $-$2.0\,dex \citep{andrievskyetal2002} while volatile elements such as carbon, nitrogen, and oxygen have near-solar abundances \citep{baschek&slettebak1988,kampetal2001,folsometal2012}.

The abundance dichotomy between refractories and volatiles suggests that accretion from a circumstellar disc plays a role in the development or maintenance of the chemical anomalies \citep{venn&lambert1990,turcotte&charbonneau1993,king1994}. The material itself does not need to be metal-weak because a variety of efficient dust-gas separation mechanisms can operate around A stars \citep{jermyn&kama2018}, allowing volatile-rich gas to be accreted onto the star without the refractory dust \citep{watersetal1992}. Suggestions for the accretion source have included material left-over from star formation \citep{holweger&stuerenburg1993}, gas from dense regions of the ISM \citep{kamp&paunzen2002}, and material ablated from hot jupiters \citep{jura2015}. Protoplanetary discs are particularly likely sources, especially since embedded planets can deplete the dust in a way that reproduces observed $\lambda$\,Boo abundances \citep{kamaetal2015,jermyn&kama2018}. VLTI and ALMA observations confirm the existence of planets embedded in the discs of some $\lambda$\,Boo stars \citep[e.g.][]{matteretal2016,fedeleetal2017,cugnoetal2019,tocietal2020}.

Numerical calculations have shown that peculiarities from selective accretion ought to persist for only $10^6$\,yr once accretion has stopped \citep{turcotte&charbonneau1993}, before particle transport processes erase the chemical abundance signature. It then follows that most $\lambda$\,Boo stars should be actively accreting. However, \citet{grayetal2017} found that $\lambda$\,Boo stars were no more likely to be observed with a debris disc at 22\,$\upmu$m than chemically normal A stars. It is also apparent that not all $\lambda$\,Boo stars are young: they are found at a wide range of main-sequence ages when placed on an HR diagram, according to either their spectroscopic $\log g$ values \citep{iliev&barzova1995} or luminosities derived from precise {\em Gaia} parallaxes \citet{murphy&paunzen2017}. 

The age range of $\lambda$\,Boo stars suggests a reservoir of material may be needed from which the star can accrete at an arbitrary age. Such a reservoir may include comets, such as the 400-Earth-mass cloud of CO-rich comets postulated to orbit the A stars HD\,21997 and 49\,Cet \citep{zuckerman&song2012}. So-called `swarms' of comets, not unlike the fragmented comet Shoemaker--Levy 9 that delivered large quantities of volatiles to Jupiter \citep{lellouchetal1997}, have been used to explain the peculiar transits of the \textit{Kepler} A star KIC\,8462852 \citep{boyajianetal2016,bodman&quillen2016}. Such bodies, sometimes called Falling Evaporating Bodies (FEBs), may be perturbed from dormant orbits by mean-motion resonances with massive planets \citep{freistetteretal2007} or encounters with nearby stars (\citealt{bailer-jones2015b}; \citealt{grayetal2017}). This comet-reservoir scenario has a pedigree in $\beta$\,Pic \citep{king&patten1992,gray&corbally2002}, a planet-host with $\lambda$\,Boo-like properties \citep{lagrangeetal2010,chengetal2016,snellen&brown2018}, FEB-like spectral absorption signatures \citep{ferletetal1987,karmannetal2001,thebault&beust2001,karmannetal2003}, and transiting exocomets \citep{ziebaetal2019}.

A solution to the $\lambda$\,Boo puzzle requires a broad approach, including particle transport models for stars and discs, and a larger and better characterised set of observations. To address the latter, \citet{murphyetal2015b} re-investigated all known and candidate $\lambda$\,Boo stars to create a homogeneous catalogue of class members, resulting in 64 bona-fide $\lambda$\,Boo stars and 45 candidates for which more observations are required for a definite classification. Since $\lambda$\,Boo stars are rare, with only 2\% of A stars belonging in the class \citep{gray&corbally1998}, further expansion of the membership list requires efficient target selection.

In \citet[][Paper I, hereafter]{grayetal2017}, we began a search for new $\lambda$\,Boo stars using GALEX photometry to target A stars with UV excesses. The $\lambda$\,Boo stars have reduced line blanketing in the UV because they are metal weak, and hence show UV excesses compared to normal stars.
We found 33 new southern $\lambda$\,Boo stars and confirmed 12 others with that approach. By modelling their spectral energy distributions, we were also able to search for infrared excesses to make an unbiased assessment of the occurrence rate of discs around $\lambda$\,Boo stars, finding the aforementioned result that discs are no more likely around $\lambda$\,Boo stars at 22\,$\upmu$m than around normal stars.

This is the second paper in the series, also focussing on the southern hemisphere (declination $< +15$\,deg). Observations of northern targets are ongoing and will be presented in future papers. In this paper, we particularly target known emission-line stars. \citet{folsometal2012} observed that many emission-line A stars were also $\lambda$\,Boo stars -- an observation compatible with the hypothesis that $\lambda$\,Boo stars are active accretors. In addition to the emission-line stars, we created a target list of metal-weak objects by examining their Str\"omgren photometry. Our target selection, observations, and spectral classification procedures are described in Sec.\,\ref{sec:method}.

We also look for infrared excesses around our targets, which might indicate the source of the accreted material if the accretion episode is recent or ongoing. We describe our SED modelling and search for infrared excesses in Sec.\,\ref{sec:IR}, and present conclusions in Sec.\,\ref{sec:conclusions}.


\section{Method}
\label{sec:method}

\subsection{Sample selection}
\label{ssec:sample}

To improve the success rate of searching for $\lambda$\,Boo stars beyond the 2\% one expects at random, we compiled a target list from several types of stars that we considered likely to yield new $\lambda$\,Boo stars. Unlike \paperi, our goal was only to find more $\lambda$\,Boo stars, and although we didn't specifically favour stars with infrared excesses, we did not actively eliminate such survey bias. A major focus in this work was emission-line A stars, but relatively few ($<$50) of these are known. Blue horizontal branch (BHB) stars are another class of rare metal-weak A stars that we considered to be promising targets, since some might have been misclassified in the literature. The target list therefore contained a mixture of emission-line stars, BHB stars, and a large number of metal-poor stars selected using Str\"{o}mgren photometry. Due to good weather and efficient observing, we added a further group of targets on the final night of the observing run, comprising A and early F stars observed in Campaign 01 of the K2 Mission. Targets were organised into groups based on how they were selected (see Table\,\ref{tab:groups}):
\begin{enumerate}
\item[Group 0.] {\em Known $\lambda$\,Boo stars.} In order to verify that the spectra were suitable for accurate classification, we obtained spectra of eleven known $\lambda$\,Boo stars, chosen according to availability on the sky at the time of observation. One of these, HD\,111164, was classified as a $\lambda$\,Boo star by \citet{abt&morrell1995}, but turned out to be a chemically normal rapid rotator.
\item[Group 1.] {\em Emission-line A stars (i).} We used the criteria search function of the SIMBAD database \citep{wengeretal2000} to select spectral types matching \mbox{``A[0-9]*e'',} where `[0-9]' represents any integer in this range, the asterisk is a wildcard of any length, and `e' is the standard notation for emission lines. These are Herbig Ae/Be stars \citep{herbig1960,hillenbrandetal1992}, the hotter analogues of T\,Tauri stars \citep{joy1945,appenzeller&mundt1989}. We expected that focussing on emission line stars would increase the efficiency of our $\lambda$\,Boo search by preferentially observing stars with circumstellar discs, or stars accreting material from an unknown source. Having a larger sample of such stars is also useful for ascertaining any link between age, accretion, and $\lambda$\,Boo peculiarity. Of the 308 stars observed, 20 stars came from this group.
\item[Group 2.] {\em Emission-line A stars (ii).} This group is phenomenologically identical to the previous group, except that the search terms were slightly modified to capture stars whose spectral types had been recorded differently. We searched for object types matching \mbox{``Em*/Ae*'' {\em and} ``spectral type = A''.}  Of the 308 stars observed, 18 stars came from this group.
\item[Group 3.] {\em Photometrically metal-weak stars.} Str\"{o}mgren photometry can be used quite efficiently to select metal-weak stars from a sample of A stars. The $m_1$ index is sensitive to metallicity, with metal-weak stars having lower values of $m_1$ than normal stars at a given $b-y$ colour (see \citealt{paunzen&gray1997}). 
We selected stars using the following criteria:
\begin{itemize}
\item[i.] $-0.015 < (b-y) < 0.30$
\item[ii.] $m_1 > 0.130 - 0.3(b-y)$
\item[iii.] $m_1 < 0.220 - 0.3(b-y)$
\item[iv.] $c_1 < 1.4 - 2.0(b-y)$
\end{itemize}
and prioritised targets with Tycho $B$ magnitudes < 10 that had not already been observed by \citet{paunzen&gray1997} or other papers in that series \citep{paunzenetal2001,paunzen2001}.
Of the 308 stars observed, 210 stars came from this group.
\item[Group 4.] {\em Blue-horizontal-branch (BHB) stars.} At classification resolution ($R\sim3000$), the spectra of BHB stars are quite similar to $\lambda$\,Boo stars. We observed seven BHB stars from \citet{macconnelletal1971} as a likely source of additional $\lambda$\,Boo stars.
\item[Group K2.] {\em Targets scheduled to be observed in Campaign 01 of the K2 Mission.} Space photometry can be beneficial to the study of $\lambda$\,Boo stars in multiple ways. For instance, there are $\lambda$\,Boo stars with exoplanets, such as HR\,8799 \citep{soummeretal2011}, so space photometry might reveal exoplanet (or exocomet) transits around $\lambda$\,Boo stars. In addition, the same photometry can be used for asteroseismology. Stellar oscillations are sensitive to metallicity, and can be used to determine whether stars are globally metal-poor or just have surface peculiarities \citep{murphyetal2013a}. 
We therefore observed some A-type stars that were scheduled to be observed in Campaign 01 of the K2 Mission \citep{howelletal2014}. This group was not selected according to spectroscopic or photometric properties, so it is numbered differently from the others. It is also not anticipated to yield a higher number of $\lambda$\,Boo stars than the 2\% expected from a random draw of field stars. Of the 308 stars observed, 42 stars came from this group, and we found one (HD\,98069) to be a $\lambda$\,Boo star. Its K2 light curve reveals it is a $\delta$\,Sct star with eight pulsation peaks exceeding 1\,mmag and a further seven exceeding 0.5\,mmag, most of which lie between 12 and 18\,d$^{-1}$. Further asteroseismic analysis is beyond the scope of this work. A TESS light-curve is also available, has similar properties, and has been analysed along with the lightcurves of all southern $\lambda$\,Boo stars by \citet{murphyetal2020b}.
\end{enumerate}

Our target list reflects our single-site, single-epoch observations (Sec.\,\ref{ssec:telescope}): only targets observable during 2014 Mar were included, corresponding roughly to right ascension in the range 75--300$^{\circ}$. Our focus on emission line stars (Groups 1 and 2) produced many new targets not already searched for $\lambda$\,Boo stars, whereas the Str\"omgren targets (Group 3) have an overlap of 21 targets with \paperi, which were observed at SAAO in 2013 and 2014. Some overlap is desirable to check for consistency between different instruments, noting of course that some targets may be spectrum variables. Because some of those 21 overlapping stars are $\lambda$\,Boo stars, they are co-discoveries. Two stars (HD\,94326 and HD\,102541) whose SAAO spectra showed $\lambda$\,Boo spectral features are classified as non-$\lambda$\,Boo metal-weak stars in this work. More spectra and an abundance analysis are desirable to confirm whether these are indeed $\lambda$\,Boo stars, and to analyse the variability in their spectra. Other than the 21 overlapping targets and the 11 in Group 0, the remainder (276) were unique to this survey.

\begin{table}
\caption{Breakdown of the target selection groups described in Sec.\,\ref{ssec:sample}, the number of stars in each group ultimately classified as $\lambda$\,Boo stars (including the two uncertain ``$\lambda$\,Boo?'' stars), the total number of targets in each group, and the percentage of $\lambda$\,Boo stars obtained by dividing the previous two columns.}
\centering
\resizebox{\columnwidth}{!}{%
\begin{tabular}{c c r r r}
\toprule
Group & Description & Num. $\lambda$\,Boo & Total & Percent \\
\midrule 
0 & Known $\lambda$\,Boo stars & 10 & 11 & 91\% \\
1 & ``A[0-9]*e'' & 4 & 20 & 20\% \\
2 & ``Em*/Ae*'' {\it and} ``A'' & 1 & 18 & 6\% \\
3 & Photometrically met wk & 16 & 210 & 8\% \\
4 & BHB stars & 2 & 7 & 29\% \\
K2 & K2 targets & 1 & 42 & 2\% \\
\bottomrule
\label{tab:groups}
\end{tabular}
}
\end{table}


\subsection{Observations}
\label{ssec:telescope}
During 2014~Mar~17--19 we obtained spectra of 308 targets with the WiFeS spectrograph \citep{dopitaetal2007} on the ANU 2.3-m telescope at Siding Spring Observatory. Our spectra were obtained in the blue-violet region in B3000 mode and have a resolution of about 2.5\,\AA/2 pixels. The WiFeS data were reduced with the {\sc pywifes} software package \citep{childressetal2014}. Due to difficulty in rectifying the spectra over the Balmer jump, we trimmed the spectra to the range 3865--4960\,\AA. The spectra thus cover the region between the blue wing of H8 and the red wing of H$\beta$. The spectra are qualitatively similar to those made from SAAO for \paperi.


\subsection{Spectral classification}
\label{ssec:class}

We classified the spectra on the MK system, which is described by \citet{gray&corbally2009}. The $\lambda$\,Boo stars are described in detail there and in \paperi, so we give only a summary here. When classifying A stars, the three main temperature criteria are (i) the strength of the \ion{Ca}{ii} K line, which rapidly increases towards later (cooler) types; (ii) the strength of the Balmer lines of hydrogen, which have a broad maximum around A2 and decrease on either side; and (iii) the metal lines, which increase in strength almost uniformly from A0 to F0. Ordinarily, all of these are absorption lines and in a normal star, all three criteria would yield the same temperature subclass. This is not the case in the $\lambda$\,Boo stars, where the metal lines are weak for a given hydrogen line type. It is the hydrogen lines that give the best estimate of the true stellar temperature, hence the spectra are usually classified with their hydrogen line type, then the luminosity class, then the K and metal line types, e.g. A7\,V\,kA2mA2~$\lambda$\,Boo. Spectral types of $\lambda$\,Boo stars having only mild peculiarity are written with the class name in parentheses: ``($\lambda$\,Boo)''. For F-type stars, the G-band becomes an important feature, and this is sometimes written prepended with a `g', e.g. F5\,V\,mF2gF5.

Each spectrum was classified by SJM and independently by at least one other author (ROG or CJC), and without knowledge of which group the target originated from. Any spectrum for which the initial classifications were found to disagree was reclassified by all three classifiers and discussed until agreement was reached on the best-fitting spectral type. The spectral types of the targets are given in Table\:\ref{tab:classes}. For explanations of notation used in spectral classes, e.g. `e' for emission and `s' for sharp-lined, see \citet{gray&corbally2009} and \citet{smithetal2011}.

\section{Stellar Parameters and Infrared Excesses}
\label{sec:IR}

Establishing the mechanisms that lead to $\lambda$\,Boo peculiarities requires a better understanding of the environments of the stars.  In particular, the accretion of dust-depleted material requires a reservoir, whose thermal emission might be detectable above the stellar luminosity in the infrared. To search for this, we constructed spectral energy distributions (SEDs) of our targets using stellar atmosphere models, against which we compared infrared fluxes from 2MASS and WISE. We followed the method from \paperi, which is summarised in this section.

\subsection{Physical Parameters and Reddening}
\label{ssec:params}

Stellar physical parameters were determined via $\chi^2$ minimisation between the observed spectra and a library of synthetic spectra computed with {\small SPECTRUM} \citep{gray&corbally1994} and {\small ATLAS9} \citep{castelli&kurucz2004}. The library grid has effective temperatures spanning 6500--25\,000\,K (having 50-K spacing up to 10\,000\,K, then 100-K spacing to 11\,500\,K, 500-K spacing to 13\,000\,K, and 1000-K spacing to 25\,000\,K), with log g = 3.3, 3.6, 4.0, and 4.2, and with metallicities of [M/H]= $+$0.5, $+$0.2, 0.0, $-$0.2, $-$0.5, $-$1.0, $-$1.5, and $-$2.0. We used the stellar spectral types (see Sec.\,\ref{ssec:class}) to estimate the intrinsic $(B-V)_0$ colours of the stars according to the relation in \paperi, making allowances for differences in stellar metallicity. 

Photometric fluxes were downloaded from IPAC.\footnote{\url{https://irsa.ipac.caltech.edu/}} We used Johnson B and V; 2MASS J, H and K; and WISE W1, W2, W3 and W4. Reddening ($E(B-V)$) was evaluated by comparing the observed $B-V$ colours with our intrinsic $(B-V)_0$ colours, and the infrared fluxes were dereddened with a combination of the Fitzpatrick reddening law \citep{fitzpatrick1999} and the mid-infrared extinction law of \citet{xueetal2016}. When Johnson B and V were unavailable, we used Tycho B$_{\rm T}$ and V$_{\rm T}$ \citep{hogetal2000} instead and followed a similar reddening procedure with a slightly different relation (\paperi) to account for the difference in zero-points of the two photometric systems \citep{bessell&murphy2012}.

\begin{figure*}
\begin{center}
\includegraphics[width=1.0\columnwidth]{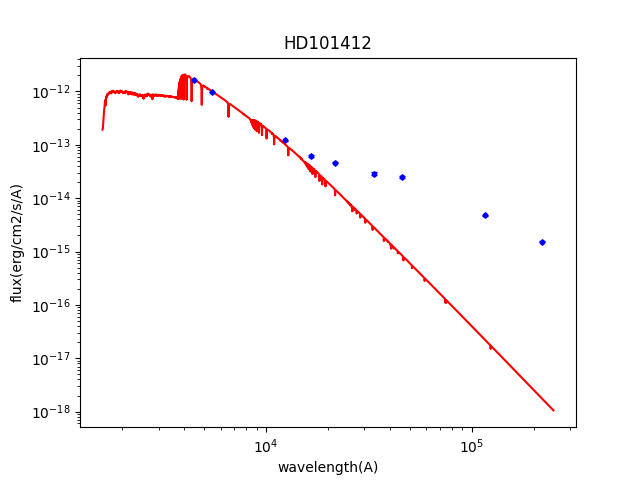}\includegraphics[width=1.0\columnwidth]{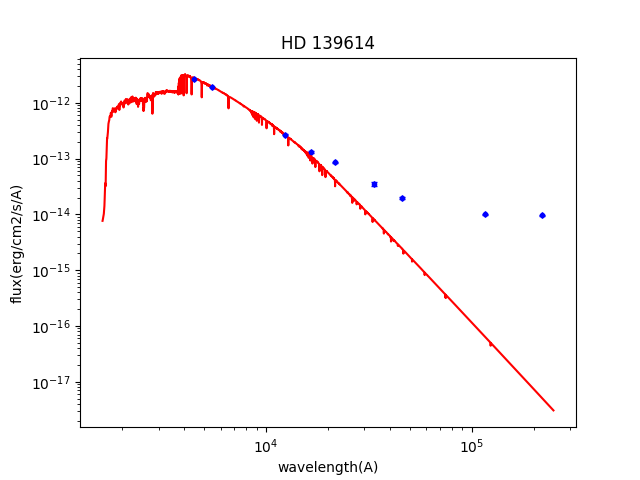}\\
\includegraphics[width=1.0\columnwidth]{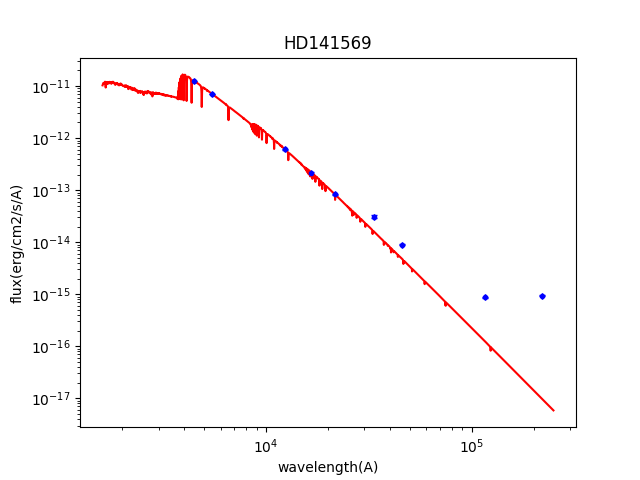}
\includegraphics[width=1.0\columnwidth]{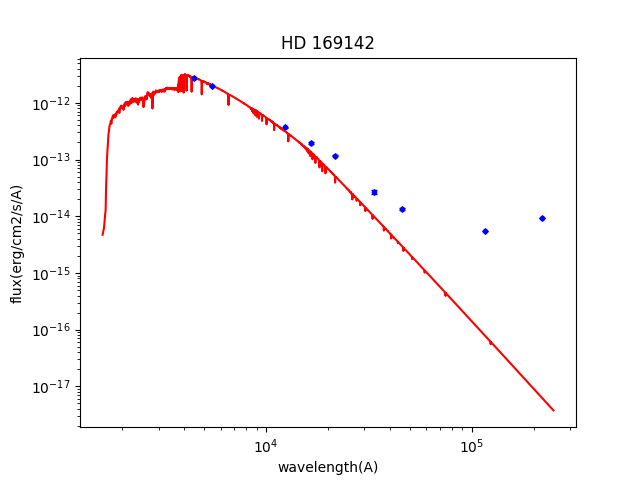}\\
\includegraphics[width=1.0\columnwidth]{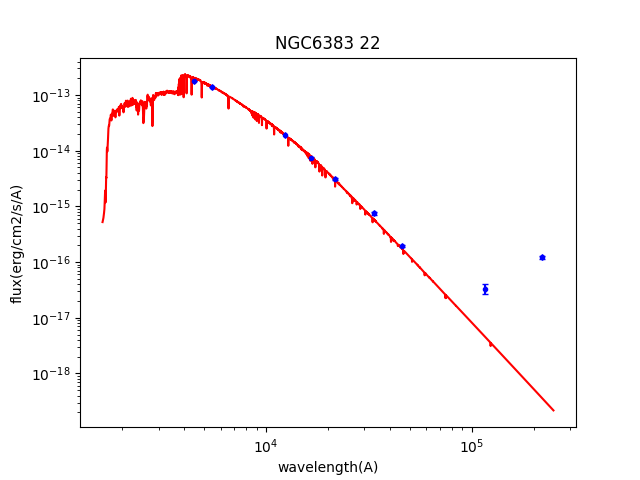}
\includegraphics[width=1.0\columnwidth]{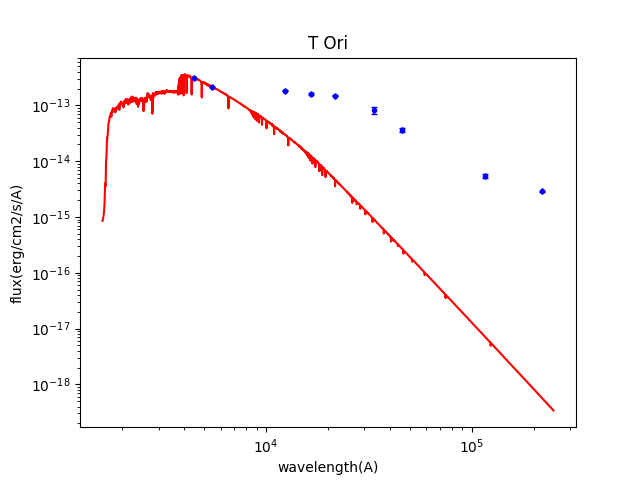}\\
\caption{Spectral energy distributions (SEDs) for the six $\lambda$\,Boo stars with strong infrared excesses ($>$10$\sigma$ in any of the WISE passbands) that probably originate from discs. Blue data points are photometric fluxes in Johnson B and V; 2MASS J, H and K; and WISE W1, W2, W3 and W4.}
\label{fig:IRx}
\end{center}
\end{figure*}

\begin{figure}
\begin{center}
\includegraphics[width=1.0\columnwidth]{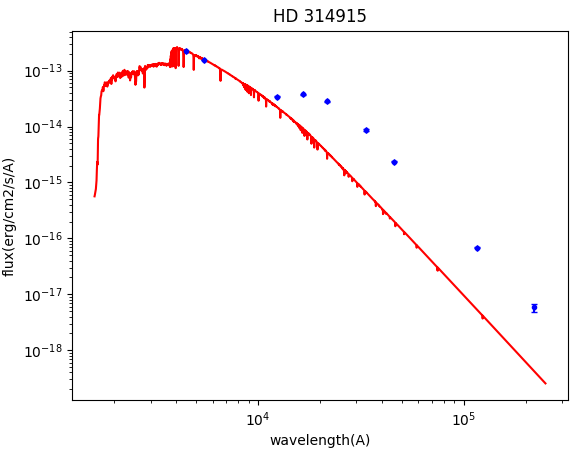}
\caption{The SED of the $\lambda$\,Boo star HD\,314915. Unlike the disc-hosts whose SEDs are shown in Fig.\,\ref{fig:IRx}, this infrared excess is more consistent with a cooler stellar companion.}
\label{fig:IRx2}
\end{center}
\end{figure}

\subsection{Infrared Excesses}
We compared the W1, W2, W3 and W4 fluxes to the synthetic spectra to identify stars with infrared excesses. We normalised the spectra to the 2MASS J band, except where there were clear excesses in the 2MASS bands, in which case spectra were normalised to the V band instead. We recorded infrared excesses (in W1--W4) in the form of a flux ratio, ($F_{\rm obs} - F_{\rm model})/F_{\rm model}$, and calculated the significance of those excesses using the recorded errors for the WISE photometry. Following \paperi, we considered infrared excesses significant at 2$\sigma$ rather than the conventional 3$\sigma$, to avoid missing potentially interesting targets for future follow-up. This is particularly important for the detection of cool discs that do not radiate strongly at wavelengths below 22\,$\upmu$m (i.e. WISE W4). Stars with excesses at $\geq$2$\sigma$ are indicated in Table\:\ref{tab:classes}, and the values and significances of the excesses are given in Table\:\ref{tab:IR}. SED parameters for stars without infrared excesses are given separately in Table\:\ref{tab:nIR}.

\begin{table*}
\centering
\caption{Infrared excesses for all stars of the sample with a $\geq$2$\sigma$ excess in one or more WISE bands. Twelve rows are shown; the full table is available online in machine-readable format. Model parameters ($T_{\rm eff}$, $\log g$, [Fe/H] and $E(B-V)$) describe the spectral energy distributions, and asterisks in the $T_{\rm eff}$ column indicate the stars for which the $V$ band rather than the $J$ band was used for normalisation. Infrared excesses (``val.'') are given as  the flux fraction in excess of the model, i.e. ($F_{\rm obs} - F_{\rm model})/F_{\rm model}$.}
\label{tab:IR}
\resizebox{\textwidth}{!}{%
\begin{tabular}{H l c H r r r r r r r r r r r r }
\toprule
Run Num & Obj. Name & Spectral Type & $\lambda$\,Boo? & $T_{\rm eff}$ & $\log g$ & [Fe/H] & $E(B$$-$$V)$ & \multicolumn{2}{c}{W1}  & \multicolumn{2}{c}{W2} & \multicolumn{2}{c}{W3} & \multicolumn{2}{c}{W4} \\
\cmidrule(lr){9-10}\cmidrule(lr){11-12}\cmidrule(lr){13-14}\cmidrule(lr){15-16}
&&&& K & & & mag & val. & $\sigma$ & val. & $\sigma$  & val. & $\sigma$ & val. & $\sigma$ \\
\midrule
366&BD-15\,1548&B7\,IIIe~He-wk&0&13000\rlap{\,$*$}&3.3&0.0&0.2&1.2&25.8&1.5&32.9&3.5&40.7&9.5&10.1\\
1177&BD-15\,4515&F2\,V\,kA4mA6~$\lambda$\,Boo&1&7000&4.2&$-$1.5&0.108&0.1&3.6&0.1&5.5&0.1&6.0&$\dots$&$\dots$\\
387&CD-37\,3833&A2\,Vn\,kA0&0&8700&4.2&$-$0.5&0.05&$\dots$&$\dots$&$\dots$&$\dots$&0.1&2.4&$\dots$&$\dots$\\
372&CD-48\,3541&A2\,Vn\,kA0mA1&0&8900\rlap{\,$*$}&4.2&$-$1.0&0.04&$\dots$&$\dots$&$\dots$&$\dots$&0.1&2.1&1.5&2.7\\
385&CD-55\,2595&B1\,Ve&0&25000\rlap{\,$*$}&4.0&0.0&0.31&0.3&12.1&0.6&20.9&1.4&19.9&$\dots$&$\dots$\\
1055&CD-59\,1764&A0.5\,V&0&9650&4.2&0.0&0.052&$\dots$&$\dots$&$\dots$&$\dots$&0.1&3.4&1.0&2.1\\
1043&CD-60\,1932&A0\,Vnn&0&9800&4.2&0.0&0.033&0.1&5.6&0.1&6.9&0.2&6.1&0.9&2.2\\
1105&CD-60\,4157&A1\,Van&0&9500&4.2&0.0&0.162&$\dots$&$\dots$&$\dots$&$\dots$&$\dots$&$\dots$&1.0&3.3\\
395&CPD-58\,3138&A1.5\,Vs&0&9200&4.2&0.0&0.06&0.1&2.8&0.0&2.4&$\dots$&$\dots$&$\dots$&$\dots$\\
1090&HD\,100380&A4\,IVs&0&8350&4.0&0.0&0.035&$\dots$&$\dots$&0.1&3.2&$\dots$&$\dots$&0.1&3.4\\
407&HD\,100453&F1\,Vn&0&7100\rlap{\,$*$}&4.2&0.0&0.0&7.1&4.0&20.9&5.3&184.4&98.7&2283.8&136.2\\
1066&HD\,101412&A3\,V(e)\,kA0.5mA0.5~($\lambda$\,Boo)&0&8500\rlap{\,$*$}&4.2&$-$1.5&0.114&9.2&16.0&28.6&18.0&208.8&108.6&888.1&90.9\\
\bottomrule
\end{tabular}
}
\end{table*}

\begin{table}
\centering
\caption{Parameters from SED fitting, for the stars without detected infrared excesses. Asterisks in the $T_{\rm eff}$ column indicate the stars for which the $V$ band rather than the $J$ band was used for normalisation. The full machine-readable table is available online.}
\label{tab:nIR}
\resizebox{\columnwidth}{!}{%
\begin{tabular}{l c H H r r r r}
\toprule
Obj. Name & Spectral Type & $\lambda$\,Boo? & Norm & $T_{\rm eff}$ & $\log g$ & [Fe/H] & $E(B$$-$$V)$ \\
& & & & K & & & mag\\
\midrule
BD+00 2757&F5 V: mF2gF5&0&J&6500&4.2&$-0.5$&0.015\\
CD-31 4428&A2 Van&0&V&8750\rlap{\,$*$}&4.2&0.0&0.05\\
CD-58 3782&A3 Van&0&J&8700&4.2&$-0.2$&0.03\\
CD-60 1956&A0.5 V&0&J&9650&4.2&0.0&0.132\\
CD-60 1986&A2 Van&0&J&9500&4.2&0.0&0.07\\
CD-60 6017&A8 IV-V&0&J&7500&4.2&0.0&0.2\\
CD-60 6021&B7 IVn&0&V&13000\rlap{\,$*$}&3.6&0.0&0.214\\
CPD-20 1613&A0.5 V kB9.5&0&J&9500&4.2&$-0.5$&0.0\\
CPD-58 3071&A3 Va&0&V&8500\rlap{\,$*$}&4.2&0.0&0.0\\
CPD-58 3106&A1.5 Vn&0&J&9200&4.2&0.0&0.08\\
HD 100237&A1 IVs&0&J&9500&3.6&0.0&0.0\\
HD 100325&A1 Va&0&J&9500&4.2&0.0&0.172\\
\bottomrule
\end{tabular}
}
\end{table}

We find that 21 of the 34 $\lambda$\,Boo stars in our sample have IR excesses. Seven of them exceed 10$\sigma$ in strength, and six of those (HD\,101412, HD\,139614, HD\,141569, HD\,169142, NGC\,6383\,22, and T\,Ori) have excesses that are larger at longer wavelengths, suggesting circumstellar discs (Fig.\,\ref{fig:IRx}). We found emission lines in the spectrum of HD\,139614, which is known to be a pre-main sequence star with a protoplanetary disc \citep{matteretal2016,carmonaetal2017,lawsetal2020}, in the less well-studied accretor HD\,101412 \citep{cowleyetal2012,scholleretal2016}, and in the cluster member T\,Ori. For HD\,141569 and HD\,169142, we found no emission in our spectra, even though HD\,141569 is known to have a Kuiper-Belt-like debris disc \citep{mendigutiaetal2017,mawetetal2017,mileyetal2018,white_ja_etal2018,bruzzoneetal2020} and HD\,169142 has a protoplanetary disc \citep{fedeleetal2017,carneyetal2018,ligietal2018,chenetal2019,grattonetal2019,maciasetal2019,tocietal2020}. For NGC\,6383\,22, our spectrum shows weak emission. Further observations of this target would be worthwhile, especially high-resolution spectroscopy in the visible for an abundance analysis, and ALMA or VLT observations for dust characterisation. The seventh target with a $>$10$\sigma$ IR excess is HD\,314915. Although this is classified as an emission-line star on SIMBAD (from \citealt{nesterovetal1995}), its SED appears to be more consistent with a cool binary companion (Fig.\,\ref{fig:IRx2}).

Table\:\ref{tab:group_IRx} shows the fraction of stars in each target selection group with infrared excesses. The K2 targets constitute the only group that is presumably unbiased with respect to infrared excess, and in that group, out of 41 normal A-type stars, 10 show excesses at $\geq2\sigma$ in one or more WISE bands.  That is a proportion of $24.4 \pm 7.7$\%. In \paperi, 18 out of 121 normal A-type stars in the Tycho sample showed excesses, giving a proportion of $14.9 \pm 3.5$\%. According to a two-tailed Z test, the resulting z-score is 1.3225, with a p value of 0.187, so those two proportions are not significantly different.  Combining the K2 and Tycho normal star samples, we find that out of a total of 162 normal A-type stars, 28 show WISE 2$\sigma$ excesses, or a proportion of $17.3 \pm 3.3$\%.  This is similar to the $20.0 \pm 10$\% observed for $\lambda$\,Boo stars in \paperi, although a larger unbiased sample of $\lambda$\,Boo stars is clearly needed before we can make any meaningful statement about whether the proportion of $\lambda$\,Boo stars with IR excesses differs from that of normal A-type stars.

\begin{table*}
\caption{Breakdown of infrared excesses among the target selection groups described in Sec.\,\ref{ssec:sample}. We give the number of $\lambda$\,Boo stars, and their percentage of the total group numbers; the number of stars with IR excesses, and their percentage of the measurable population (i.e. group members where we could construct and evaluate SEDs for IR excesses); and the number of $\lambda$\,Boo stars with IR excesses as a percentage of the number of $\lambda$\,Boo stars in that group.} 
\centering
\begin{tabular}{c c r r r r r r r}
\toprule
Group & Description & Total & \multicolumn{2}{c}{$\lambda$\,Boo} & \multicolumn{2}{c}{IR excess} &  \multicolumn{2}{c}{$\lambda$\,Boo + IR} \\
\cmidrule(lr){4-5} \cmidrule(lr){6-7} \cmidrule(lr){8-9}
& & stars & Num. & \% & Num. & \% of group & Num. & \% of $\lambda$\,Boo \\
\midrule 
0 & Known $\lambda$\,Boo stars & 11 & 10 & 91 & 4 & 40 & 4 & 40 \\
1 & ``A[0-9]*e'' & 20 & 4 & 20 & 17 & 85 & 4 & 100\\
2 & ``Em*/Ae*'' {\it and} ``A'' & 18 & 1 & 6 & 16 & 89 & 1 & 100 \\
3 & Photometrically metal weak & 210 & 16 & 8 & 105 & 50 & 10 & 63 \\
4 & BHB stars & 7 & 2 & 29 & 4 & 57 & 1 & 50 \\
K2 & K2 targets & 42 & 1 & 2 & 11 & 26 & 1 & 100 \\
\bottomrule
\label{tab:group_IRx}
\end{tabular}
\end{table*}

\subsection{Luminosities}

To position the $\lambda$\,Boo stars in our sample on the HR diagram, we determined their luminosities. We followed the methodology of \citet{murphyetal2019} and \citet{heyetal2019}, except that we used the Johnson V band rather than SDSS g. Bolometric luminosities were calculated via absolute magnitudes using standard formulae:
\begin{equation}
M_V = m_V - 5(\log d -1) -A_V,
\label{eq:absmag}
\end{equation}
and
\begin{equation}
\log L_{\rm bol}/L_{\odot} = -(M_V + BC - M_{{\rm bol}, \odot})/2.5.
\label{eq:logL}
\end{equation}
The apparent V magnitudes, $m_V$, are those in Table\:\ref{tab:classes}, which are taken from the SIMBAD database with an assumed uncertainty of 0.02\,mag. The V-band extinctions, $A_V$, were taken as $3.1E(B-V)$, using the $E(B-V)$ values determined in Sec.\,\ref{ssec:params}. Bolometric corrections, BC, were computed via grid interpolation, taking the observed $T_{\rm eff}$, $\log g$, and [Fe/H] from SED fitting (Sec.\,\ref{ssec:params}) and uncertainties of 250\,K, 0.5\,dex and 0.25\,dex, respectively. These correspond to approximately 0.15\,mag, 0.02\,mag and 0.025\,mag of uncertainty in the BC, which we combined in quadrature. We adopted a bolometric magnitude for the Sun, $M_{{\rm bol}, \odot}$, of 4.74 \citep{mamajeketal2015b}. Distances were calculated using Gaia DR2 parallaxes \citep{gaiacollaboration2018a}, their uncertainties, and the length-scale model of \citet{bailer-jonesetal2018}. To determine luminosities with uncertainties, for each star we generated 10\,000 distance samples  that we fed into a Monte Carlo process using equations\:\ref{eq:absmag} and \ref{eq:logL}, and took the median and standard deviation of the resulting distribution.

\begin{figure}
\begin{center}
\includegraphics[width=1.0\columnwidth]{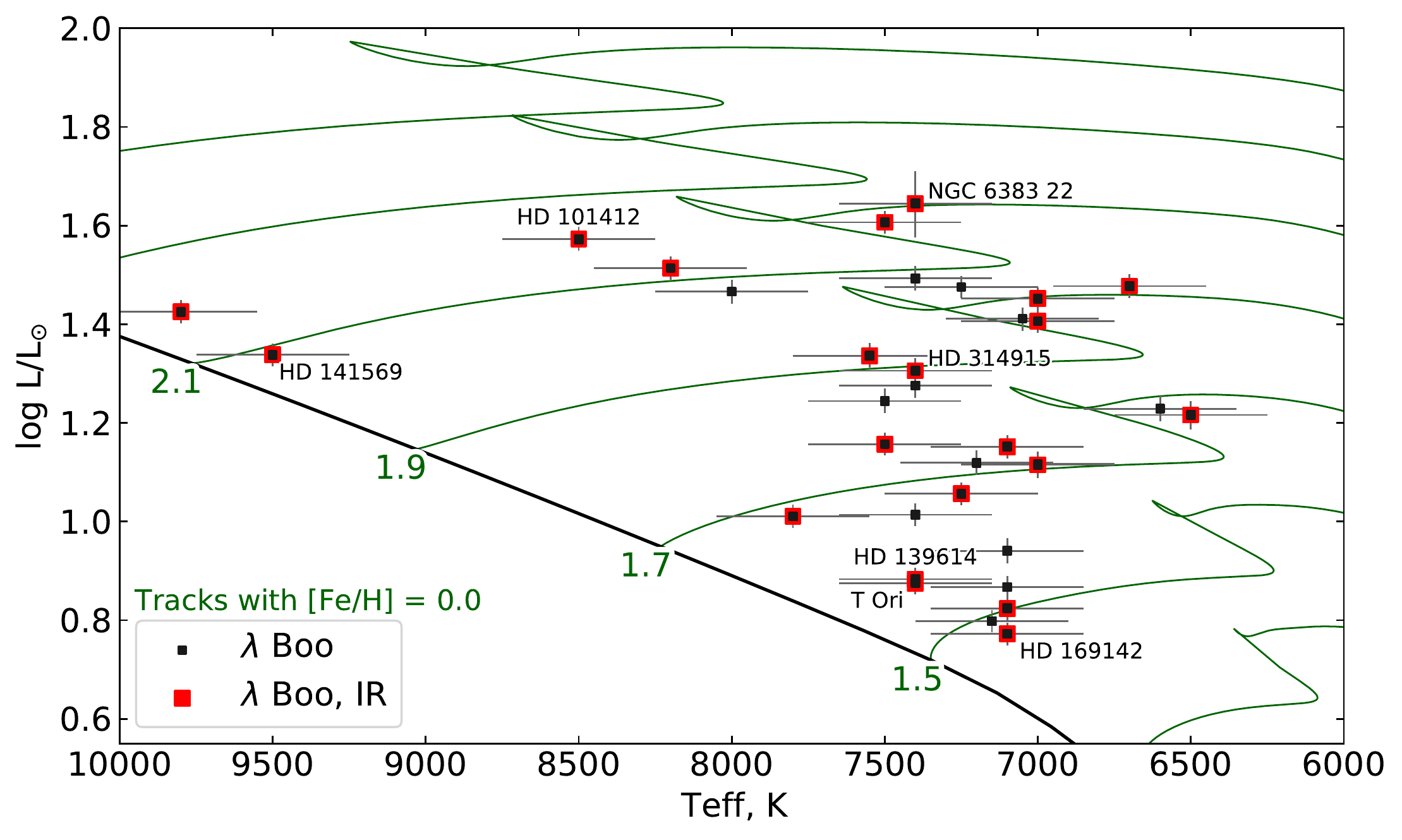}
\caption{HR diagram of the 34 $\lambda$\,Boo stars. The 21 stars with infrared excesses are highlighted with red boxes and the seven stars with $>10\sigma$ excesses are labelled (see also Fig.\,\ref{fig:IRx} and \ref{fig:IRx2}). Evolutionary tracks of solar metallicity from \citet[][green lines]{murphyetal2019} are shown at intervals of 0.2\,M$_{\odot}$.}
\label{fig:HRD}
\end{center}
\end{figure}

Using these luminosities together with the effective temperatures from SED fitting, we plot the $\lambda$\,Boo stars in an HR diagram in Fig.\,\ref{fig:HRD}. The $\lambda$\,Boo stars with infrared excesses are highlighted, some of which clearly lie near the terminal-age main sequence. This confirms earlier results \citep{paunzenetal2002b,paunzenetal2014,murphy&paunzen2017,grayetal2017}, that the $\lambda$\,Boo stars have a range of main-sequence ages. There is no apparent preference towards the ZAMS, even among the $\lambda$\,Boo stars with infrared excesses that are presumably attributable to discs.

\section{Conclusions}
\label{sec:conclusions}

The curation of a large and well defined sample of $\lambda$\,Boo stars is important for understanding the accretion environments and particle transport processes affecting A type stars more broadly. We have classified 308 stars on the MK system and discovered or co-discovered 24 new $\lambda$\,Boo stars, including two that require high-resolution spectroscopy for an abundance analysis to confirm their membership in the class. These represent a 17\% increase in the number of known $\lambda$\,Boo stars, adding to the 64 in the \citet{murphyetal2015b} catalogue and the 45 in \paperi, after accounting for overlap and revised spectral types. Our revision of 11 known $\lambda$\,Boo stars revealed that one is a chemically normal rapid rotator. This one misclassified target suggests that abundance analyses would be valuable to confirm $\lambda$\,Boo stars. 

The fraction of field A stars that are $\lambda$\,Boo is known to be approximately 2\%, whereas stars identified photometrically as being metal-weak yielded a relatively high fraction of $\lambda$\,Boo stars (8\%).
We estimate that roughly half of all stars with Str\"omgren photometry and meeting our metal-weak criteria (Sec\,\ref{ssec:sample}) have now been searched for $\lambda$\,Boo stars, but with strong bias towards higher completeness in the southern hemisphere; the northern sky is comparatively underexplored, and will be the subject of future work, along with a refinement of those selection criteria to improve search efficiency.
Our observations of 38 emission-line stars yielded 5 new $\lambda$\,Boo stars (13\%). Emission-line stars are a relatively untapped source, since our search only used emission-line objects with known spectral types. Using further SIMBAD criteria searches, we find 1126 emission-line objects without spectral types but with the correct $B-V$ colours ($-0.05$ to $0.4$) to be potential $\lambda$\,Boo stars. These should be high priority targets for future searches for $\lambda$\,Boo stars.

We collated fluxes in nine passbands to model the spectral energy distributions of all targets to look for infrared excesses. Unsurprisingly, infrared excesses were highly prevalent among the emission-line stars, including all those that are $\lambda$\,Boo stars. We also calculated stellar luminosities to plot the $\lambda$\,Boo stars on the HR diagram, confirming that not all $\lambda$\,Boo stars are young: even those that have infrared excesses are found at a variety of main-sequence ages.



\section*{Acknowledgements}
The authors thank the reviewer, Fr\'ed\'eric Royer, whose comments improved this manuscript. This work was supported by the Australian Research Council through the award DE180101104. This research has made use of the NASA/IPAC Infrared Science Archive, which is funded by the National Aeronautics and Space Administration and operated by the California Institute of Technology, It also made use of the SIMBAD database,
operated at CDS, Strasbourg, France. We used Lightkurve, a Python package for Kepler and TESS data analysis \citep{lightkurvecollaboration2018}, and {\sc spectrum} for creating synthetic spectra \citep{gray1999}.

\section*{Data Availability}
Table\,\ref{tab:classes}, which appears in full in the appendix, is also available online. Tables\,\ref{tab:IR} and \ref{tab:nIR}, which contain stars with and without infrared excesses, respectively, are each shown for twelve rows in this paper and are available in full in machine-readable format online. The stellar spectra are available from the corresponding author upon reasonable request.



\bibliographystyle{mnras}
\interlinepenalty=10000
\bibliography{sjm_bibliography} 
\bsp	

\appendix

\section{Spectral classes for the program stars}

\clearpage

\begin{table}
\centering
\caption{Spectral classes for the program stars. We give the target group of each star (Sec.\,\ref{ssec:sample}). Comments on the spectra are recorded as endnotes. Infrared excesses are denoted with `1' in the final column.}
\label{tab:classes}
\begin{savenotes}
\resizebox{\columnwidth}{!}{%
\begin{tabular}{H l r c c l c}
\toprule
Run Num & Obj Name & \hspace{-2mm}V mag & \hspace{-2mm}Group\hspace{-2mm} & Class & \hspace{-1mm}Note & IR\\
\midrule
1019 & HD\,28490 & 9.53 & 3 & F0\,V(n)\,kA5mA5~($\lambda$\,Boo) & \tablefootnote{A mild $\lambda$\,Boo star.} &  \\
1020 & HD\,29650 & 9.66 & 3 & A3\,IV-Vs &  & 1 \\
1010 & HD\,30335 & 9.67 & 2 & A4\,IV Sr &  &  \\
1021 & HD\,32725 & 9.52 & 3 & F3\,V Sr & \tablefootnote{Especially clear enhancement of \specline{Sr}{ii}{4215}. Might be an early Ba dwarf.} & 1 \\
1023 & HD\,33901 & 9.52 & 3 & A7\,III &  &  \\
1011 & HD\,35343 & 10.25 & 1 & Be3 & \tablefootnote{Strong emission in H lines and \ion{Fe}{ii} lines} & 1 \\
1026 & HD\,35793 & 9.77 & 3 & A2\,Vs &  & 1 \\
365 & HD\,36121 & 8.98 & 1 & kA6hA8mA8\,III~Sr & \tablefootnote{Mild Am peculiarity.} &  \\
1031 & HD\,36866 & 9.52 & 3 & A3\,Vas &  & 1 \\
1030 & HD\,36899 & 9.8 & 3 & A1\,V &  &  \\
1032 & HD\,36955 & 9.58 & 3 & A7\,Vp~SrEu &  &  \\
1039 & HD\,37091 & 9.82 & 3 & A2.5\,V &  & 1 \\
1038 & HD\,37258 & 9.61 & 3 & A3\,V~shell & \tablefootnote{Shell core in H\,$\beta$ and \specline{Fe}{ii}{4233}.} & 1 \\
364 & HD\,37357 & 8.85 & 1 & A3\,Van\,kA1 &  & 1 \\
1033 & HD\,37412 & 9.76 & 3 & A2.5\,Vs &  & 1 \\
1037 & HD\,37455 & 9.6 & 3 & A3\,Vb &  & 1 \\
1036 & HD\,37469 & 9.58 & 3 & B9\,Vp~Si-Sr &  & 1 \\
1013 & HD\,40632 & 9.15 & 1 & B9\,IV~shell & \tablefootnote{Strong metallic-line spectrum, similar to F0\,III. The \specline{Fe}{ii}{4233} line is strong and the hydrogen line cores are deep.} & 1 \\
1008 & HD\,44351 & 8.25 & 1 & F5\,V~composite & \tablefootnote{The K line is broad and shallow, while metal lines are of mixed strengths.} & 1 \\
1018 & HD\,46390 & 10.08 & 2 & B7\,IV-Ve & \tablefootnote{Emission reversal in H\,$\beta$. H\,$\gamma$ and H\,$\delta$ partially filled with emission.} & 1 \\
371 & HD\,50937 & 9.61 & 3 & A2\,IVn &  &  \\
368 & HD\,51480 & 6.93 & 2 & B3/5\,Ibe & \tablefootnote{P Cygni profile} & 1 \\
369 & HD\,55637 & 9.65 & 2 & B6\,IV & \tablefootnote{Classified in Simbad as emission line Ap Si. No sign of emission or increased abundance of Si.} & 1 \\
370 & HD\,59000 & 9.57 & 3 & A8\,IV/V &  &  \\
386 & HD\,62752 & 8.11 & 3 & B9\,Vp~SiEuSr & \tablefootnote{Very peculiar.} &  \\
382 & HD\,63524 & 8.8 & 2 & B6.5\,Vn &  &  \\
1042 & HD\,63562 & 9.69 & 3 & A1\,V &  & 1 \\
1058 & HD\,66318 & 9.56 & 3 & A2:\,IV:p~SiSrCr & \tablefootnote{Very peculiar. Temperature type very uncertain. H lines do not fit well at any spectral type.} &  \\
1059 & HD\,67658 & 9.76 & 3 & A4\,IV &  &  \\
383 & HD\,68695 & 9.87 & 1 & A3\,Vbe\,kA0mA0.5 & \tablefootnote{Emission in the core of H\,$\beta$. \specline{Mg}{ii}{4481} is normal. Not a $\lambda$\,Boo star.} & 1 \\
388 & HD\,75185 & 9.82 & 3 & A2\,IV-n. &  &  \\
384 & HD\,79066 & 6.34 & 1 & F1\,Vn & \tablefootnote{Rapid rotation gives the impression of metal weakness unless comparing against high $v\sin i$ standards.} &  \\
389 & HD\,80692 & 9.69 & 3 & F0\,V~+~Ae~composite & \tablefootnote{H\,$\beta$ has broad wings and a deep narrow core, suggesting shell, emission, or composite.  K-line about A1.  H$\gamma$ is F0\,V.  Note that Simbad has this as an eclipsing binary.} &  \\
1218 & HD\,83041 & 8.79 & 4 & F1.5\,V\,kA3mA3~$\lambda$\,Boo &  &  \\
390 & HD\,83798 & 9.58 & 3 & A5\,IVnn &  &  \\
391 & HD\,85337 & 9.61 & 3 & hA9\,Vn\,kA6mA6 & \tablefootnote{More rapidly rotating than the Vn standards. Metal weak even after considering rotation.} &  \\
358 & HD\,87271 & 7.13 & 0 & A8\,V\,kA0mA0.5~$\lambda$\,Boo &  & 1 \\
1060 & HD\,87593 & 9.62 & 3 & A2\,Vas &  &  \\
1067 & HD\,88554 & 9.32 & 3 & F3\,V\,kA6mA8~($\lambda$\,Boo) &  & 1 \\
1068 & HD\,88976 & 6.54 & 3 & A2\,IV-V &  & 1 \\
1062 & HD\,89234 & 9.79 & 3 & A0.5\,IV &  & 1 \\
1069 & HD\,91839 & 8.39 & 3 & A3\,Vas~(met wk A2) & \tablefootnote{Not a $\lambda$\,Boo star.} & 1 \\
1063 & HD\,92251 & 9.81 & 3 & A0.5\,Vas &  &  \\
1064 & HD\,93264 & 9.54 & 3 & kA1.5hA2mA3\,IV-V & \tablefootnote{A mild Am star.} &  \\
1065 & HD\,93746 & 9.52 & 3 & F3\,V &  &  \\
1071 & HD\,93925 & 9.24 & 3 & A0\,II-IIIp~Eu &  & 1 \\
1072 & HD\,94326 & 7.76 & 3 & A6\,III\,kA5 &  & 1 \\
1073 & HD\,95883 & 7.33 & 3 & A1\,Van & \tablefootnote{Shallow H cores.} & 1 \\
402 & HD\,96040 & 9.97 & 3 & B9\,IIIp~Si &  & 1 \\
400 & HD\,96089 & 9.78 & 3 & A1\,Van &  & 1 \\
396 & HD\,96091 & 9.57 & 3 & A0.5\,Van &  &  \\
401 & HD\,96157 & 9.82 & 3 & A2\,Van &  &  \\
397 & HD\,96192 & 9.66 & 3 & A3\,Van~kA1 &  & 1 \\
403 & HD\,96304 & 9.54 & 3 & A0.5\,Van &  & 1 \\
398 & HD\,96341 & 9.53 & 3 & A0.5\,Van &  & 1 \\
399 & HD\,96386 & 9.83 & 3 & A2\,IV-V &  & 1 \\
392 & HD\,96430 & 8.49 & 2 & B6\,IV/Ve & \tablefootnote{Emission in core of H\,$\beta$, possibly causing other H lines to be shallower.} & 1 \\
1074 & HD\,96493 & 8.5 & 3 & A0.5\,III~shell &  & 1 \\
\bottomrule
\end{tabular}
}
\end{savenotes}
\end{table}

\setcounter{table}{0}
\renewcommand{\thetable}{A\arabic{table}}
\begin{table}
\centering
\caption{{\bf continued.} Spectral classes for the program stars.}
\begin{savenotes}
\resizebox{\columnwidth}{!}{%
\begin{tabular}{H l r c c l c}
\toprule
Run Num & Obj Name & \hspace{-2mm}V mag & \hspace{-2mm}Group\hspace{-2mm} & Class & \hspace{-1mm}Note & IR\\
\midrule
405 & HD\,96667 & 9.58 & 3 & A1\,Van &  &  \\
404 & HD\,96773 & 9.69 & 3 & A1\,Van &  & 1 \\
1270 & HD\,97230 & 8.62 & K2 & A7\,IV~(met str F2) & \tablefootnote{Not an Am star since the K line is strong, too.} &  \\
1234 & HD\,97340 & 8.15 & K2 & A9\,V\,mA6 &  &  \\
1269 & HD\,97373 & 8.67 & K2 & A4\,IVn &  &  \\
1075 & HD\,97528 & 7.31 & 3 & A2\,IIIe~shell &  & 1 \\
1268 & HD\,97678 & 8.67 & K2 & F2\,Vs &  &  \\
1272 & HD\,97859 & 9.35 & K2 & B8\,IVp~Si &  &  \\
1257 & HD\,97891 & 8.33 & K2 & F5.5\,V &  &  \\
1271 & HD\,97916 & 9.2 & K2 & F5.5\,V\,gF2.5kF2:mA6 &  & 1 \\
1287 & HD\,97991 & 7.41 & K2 & B1\,V &  & 1 \\
1232 & HD\,98069 & 8.16 & K2 & A9\,V\,kA2mA2~($\lambda$\,Boo) & \tablefootnote{A mild $\lambda$\,Boo star, with fluted H\,$\gamma$ lines} & 1 \\
1233 & HD\,98563 & 8.27 & K2 & F7\,V &  &  \\
1266 & HD\,98575 & 9.12 & K2 & kA5hA9mF3\,III & \tablefootnote{A $\rho$\,Pup star.} &  \\
1253 & HD\,98632 & 7.57 & K2 & F4\,Vs\,mF1 &  & 1 \\
1265 & HD\,98645 & 8.8 & K2 & F1\,Vs &  &  \\
1264 & HD\,98686 & 7.65 & K2 & A8\,Vnn &  &  \\
1254 & HD\,98711 & 8.07 & K2 & F6\,IV-V &  &  \\
1255 & HD\,98914 & 8.08 & K2 & F5.5\,V &  &  \\
1297 & HD\,99210 & 6.74 & K2 & kA8hA9mF2\,III: & \tablefootnote{Mild Am, with anomalous luminosity effect.} & 1 \\
1273 & HD\,99304 & 8.58 & K2 & F5\,IV &  &  \\
1267 & HD\,99776 & 9.18 & K2 & A2.5\,Vas &  &  \\
1246 & HD\,100237 & 7.34 & K2 & A1\,IVs &  &  \\
1089 & HD\,100325 & 9.28 & 3 & A1\,Va &  &  \\
1090 & HD\,100380 & 6.78 & 3 & A4\,IVs &  & 1 \\
1279 & HD\,100415 & 9.06 & K2 & kA6hA8mF1~(IV-III) & \tablefootnote{Marginal Am star. Anomalous luminosity effect evident.} &  \\
1252 & HD\,100417 & 8.03 & K2 & A1\,Vas &  &  \\
407 & HD\,100453 & 7.79 & 1 & F1\,Vn &  & 1 \\
1251 & HD\,100630 & 7.88 & K2 & A1.5\,Va &  &  \\
1276 & HD\,100762 & 9.32 & K2 & F4\,Vs &  &  \\
1235 & HD\,100995 & 8.09 & K2 & F4.5\,V &  &  \\
1258 & HD\,101196 & 8.5 & K2 & F4\,Vs &  &  \\
439 & HD\,101268 & 9.55 & 3 & F1\,Vs\,kA8mA6 & \tablefootnote{Metal weak overall, but less so in K line. Weakness of \specline{Ca}{i}{4226} and the difference between the K and m types suggest this is not a $\lambda$\,Boo star, despite weak \specline{Mg}{ii}{4481} line. May be composite.} &  \\
1066 & HD\,101412 & 9.29 & 3 & A3\,V(e)\,kA0.5mA0.5 ($\lambda$\,Boo)  & \tablefootnote{Slight emission notch in H\,$\beta$. A mild $\lambda$\,Boo star.} & 1 \\
1248 & HD\,101784 & 7.54 & K2 & A0\,Vas &  &  \\
1263 & HD\,101846 & 7.87 & K2 & A4\,Vs &  & 1 \\
1247 & HD\,101969 & 7.54 & K2 & F4\,V &  & 1 \\
1250 & HD\,102059 & 7.76 & K2 & F4\,Vs &  & 1 \\
1277 & HD\,102083 & 8.58 & K2 & F0\,V\,mA7 &  &  \\
1230 & HD\,102284 & 8.54 & K2 & F3\,IVs &  & 1 \\
1249 & HD\,102331 & 7.57 & K2 & F4\,Vs &  &  \\
1231 & HD\,102332 & 8.51 & K2 & F4\,IVs &  & 1 \\
1281 & HD\,102431 & 8.95 & K2 & F5.5\,V &  &  \\
1091 & HD\,102519 & 8.66 & 3 & A1\,IVn &  & 1 \\
1092 & HD\,102541 & 7.94 & 3 & hA9\,V\,kA5mA6 & \tablefootnote{Not a $\lambda$\,Boo star -- \specline{Mg}{ii}{4481} is not additionally weak. \specline{Fe}{i}{4046} is peculiarly strong in absorption.} &  \\
1259 & HD\,102731 & 8.49 & K2 & A6\,IVs &  &  \\
1283 & HD\,103547 & 9.38 & K2 & F1\,Vs\,mF2.5 &  & 1 \\
1289 & HD\,103631 & 8.53 & K2 & F8\,IV &  &  \\
1261 & HD\,103695 & 8.52 & K2 & A6\,V &  &  \\
1260 & HD\,104367 & 7.78 & K2 & F6\,IV &  &  \\
1093 & HD\,104446 & 9.05 & 3 & A1\,IVs &  & 1 \\
1285 & HD\,104624 & 9.13 & K2 & A4\,V &  &  \\
424 & HD\,104650 & 9.78 & 3 & A1\,Vas\,mA2 & \tablefootnote{H lines fit best at A1\,V, and are too narrow for A2. A2\,IV is not as good a fit as A1\,V. Metal lines and K line are strong for A1, being $\sim$A2.} & 1 \\
1094 & HD\,104697 & 9.46 & 3 & A1\,Va &  & 1 \\
1095 & HD\,105015 & 8.64 & 3 & A1\,Vas &  &  \\
1096 & HD\,105194 & 9.32 & 3 & hA0.5\,Van\,kB9.5mA0 & \tablefootnote{Not a $\lambda$\,Boo star, just marginally metal weak.} &  \\
1097 & HD\,105209 & 8.67 & 3 & A2\,IVn &  & 1 \\
1098 & HD\,105232 & 8.66 & 3 & A2\,Vs &  &  \\
1099 & HD\,105649 & 9.83 & 3 & A2\,IV-V &  & 1 \\
1126 & HD\,106373 & 8.9 & 4 & F5:\,Ia:\,kA3mA3 & \tablefootnote{Very peculiar spectrum.  Very weak H lines, which can be approximately fitted by a B7 supergiant, or a mid-late F supergiant.  In the case of the latter, the star is profoundly metal weak. Most likely a pop II star, possibly a high-latitude F supergiant, although the spectrum is peculiar even for that class.} & 1 \\
1100 & HD\,106961 & 8.93 & 3 & A0\,Vann &  &  \\
1101 & HD\,107049 & 9.36 & 3 & A0\,IV+ &  & 1 \\
\bottomrule
\end{tabular}
}
\end{savenotes}
\end{table}

\setcounter{table}{0}
\renewcommand{\thetable}{A\arabic{table}}
\begin{table}
\centering
\caption{{\bf continued.} Spectral classes for the program stars.}
\begin{savenotes}
\resizebox{\columnwidth}{!}{%
\begin{tabular}{H l r c c l c}
\toprule
Run Num & Obj Name & \hspace{-2mm}V mag & \hspace{-2mm}Group\hspace{-2mm} & Class & \hspace{-1mm}Note & IR\\
\midrule
1102 & HD\,107096 & 9.37 & 3 & A0:\,III:p~Eu & \tablefootnote{Very chemically peculiar star. H lines much deeper than A0\,III, but wings agree.} & 1 \\
434 & HD\,107127 & 9.57 & 3 & A1\,IV &  &  \\
1103 & HD\,107233 & 7.36 & 0 & F0\,V\,kA3mA2~$\lambda$\,Boo &  &  \\
1127 & HD\,107369 & 9.6 & 4 & A3\,IIp & \tablefootnote{Luminosity criteria (e.g. $\lambda\lambda$4172--8) do not agree with the hydrogen line type of A3\,II, but otherwise a good match.} & 1 \\
1104 & HD\,107483 & 9.3 & 3 & A1\,IV-III &  &  \\
425 & HD\,107878 & 9.71 & 3 & A9\,V\,mA2 & \tablefootnote{Not a $\lambda$\,Boo star, maybe pop II or composite A + F.} &  \\
1106 & HD\,108417 & 8.98 & 3 & A2\,V &  & 1 \\
1107 & HD\,108889 & 8.86 & 3 & A1\,IVs &  & 1 \\
1108 & HD\,108925 & 6.45 & 3 & A3\,IVn &  &  \\
1109 & HD\,109065 & 8.16 & 3 & A1\,IVn &  &  \\
1110 & HD\,109183 & 9.1 & 3 & A1\,Va+s &  &  \\
1112 & HD\,109435 & 8.99 & 3 & A7\,IV &  & 1 \\
1111 & HD\,109443 & 9.25 & 3 & F3\,V\,kF1mF1 &  &  \\
1113 & HD\,109517 & 8.77 & 3 & kA0hA0mA1\,IV-V &  & 1 \\
1114 & HD\,109738 & 8.3 & 0 & hA9\,Vn\,kA0mA0~$\lambda$\,Boo & \tablefootnote{Definite $\lambda$\,Boo star. \specline{Mg}{ii}{4481} is extremely weak.} &  \\
433 & HD\,109791 & 9.75 & 3 & A0\,II-IIIp SrSiEu~shell? & \tablefootnote{Highly peculiar star. H lines, especially. H\,$\beta$ is deeper than the standard, may be shell absorption.} &  \\
1115 & HD\,109800 & 8.84 & 3 & A1\,IV+s &  & 1 \\
1116 & HD\,109808 & 7.13 & 3 & A3\,V & \tablefootnote{Slightly weak K line (A2).} &  \\
1117 & HD\,109886 & 8.61 & 3 & A1\,Van & \tablefootnote{Excellent match to standard, HR\,2324.} &  \\
1118 & HD\,110640 & 9.0 & 3 & A2\,Vas &  &  \\
1119 & HD\,111105 & 7.25 & 3 & A3\,IVn &  & 1 \\
1139 & HD\,111164 & 6.09 & 0 & A4\,V(n) &  &  \\
432 & HD\,111209 & 9.62 & 3 & A4\,Vn &  & 1 \\
1120 & HD\,111438 & 9.18 & 3 & A1.5\,Vas &  &  \\
1121 & HD\,111439 & 8.87 & 3 & A3\,IV-Vs &  &  \\
1122 & HD\,111786 & 6.14 & 0 & F0\,Vs\,kA1mA1~$\lambda$\,Boo & \tablefootnote{Extreme $\lambda$\,Boo star.} & 1 \\
1123 & HD\,112938 & 8.16 & 3 & A2\,IV-V &  & 1 \\
1124 & HD\,113199 & 8.81 & 3 & A2\,V &  & 1 \\
1128 & HD\,113660 & 9.32 & 3 & A6\,IVs &  & 1 \\
1129 & HD\,113807 & 7.56 & 3 & A2\,Vas &  &  \\
1130 & HD\,114477 & 8.4 & 3 & A1.5\,IVs &  &  \\
1131 & HD\,114738 & 7.81 & 3 & A1\,IV-V &  &  \\
1132 & HD\,114836 & 8.74 & 3 & A0.5\,Van & \tablefootnote{H lines are truly halfway between A0\,Van and A1\,Van. Metal lines are consistent with this.} &  \\
1133 & HD\,115843 & 9.34 & 3 & A1\,IVs &  &  \\
1134 & HD\,116137 & 9.09 & 3 & B9.5\,Van &  &  \\
1135 & HD\,119561 & 9.79 & 3 & A1\,Van &  &  \\
1136 & HD\,119896 & 8.22 & 3 & F5\,Vs\,kA5mA5~$\lambda$\,Boo &  & 1 \\
1137 & HD\,120122 & 9.11 & 3 & F1\,Vs\,kA6mA6~($\lambda$\,Boo) &  &  \\
1141 & HD\,120873 & 9.36 & 3 & A1\,Van &  &  \\
1142 & HD\,121875 & 9.26 & 3 & A2\,IV-Vn &  &  \\
1140 & HD\,122264 & 9.59 & 3 & A0\,IIIp~EuSr &  &  \\
1143 & HD\,122757 & 8.8 & 3 & A3\,Va+s &  &  \\
414 & HD\,123960 & 9.75 & 3 & B9\,IIIp~Si &  & 1 \\
1144 & HD\,124228 & 7.86 & 3 & A3\,IV+s &  &  \\
415 & HD\,124878 & 9.54 & 3 & B8\,V~He-wk~+~A & \tablefootnote{Composite spectrum.} &  \\
1145 & HD\,126164 & 9.41 & 3 & A0\,Vbn &  &  \\
1146 & HD\,126627 & 9.0 & 3 & F1\,V\,kA5mA5~($\lambda$\,Boo) &  & 1 \\
1147 & HD\,127659 & 9.31 & 3 & F2\,V\,kA3mA4~$\lambda$\,Boo &  & 1 \\
450 & HD\,128336 & 9.08 & 3 & F4\,V\,kA2mA2~$\lambda$\,Boo? & \tablefootnote{Late $\lambda$\,Boo candidate. Needs abundance analysis to decide.} &  \\
413 & HD\,129389 & 9.68 & 3 & A0\,Van &  & 1 \\
1176 & HD\,130156 & 9.35 & 4 & kA6hF1mF3\,(II) & \tablefootnote{H lines are not consistent, F0/3.} &  \\
1148 & HD\,133800 & 6.4 & 3 & A6\,V\,kA0.5mA0.5~$\lambda$\,Boo &  & 1 \\
1149 & HD\,134685 & 7.67 & 3 & A1\,V &  & 1 \\
1150 & HD\,135284 & 9.23 & 3 & A3\,IV+s &  &  \\
448 & HD\,136463 & 9.54 & 3 & F1\,V\,kF1mA7 & \tablefootnote{Difference in K and M type argues against a $\lambda$\,Boo classification.} &  \\
1161 & HD\,137128 & 7.1 & 3 & A3\,IV-V &  &  \\
1151 & HD\,138274 & 8.8 & 3 & A1\,IVs &  &  \\
1159 & HD\,138753 & 8.54 & 3 & A0.5\,Van &  &  \\
431 & HD\,138921 & 9.72 & 3 & A9\,Vn kA4mA4~$\lambda$\,Boo & \tablefootnote{Hydrogen line is at A9, when comparing to the A9\,Vn standard.} &  \\
1152 & HD\,139612 & 9.27 & 3 & A0\,IVs &  &  \\
453 & HD\,139614 & 8.24 & 1 & A9\,Vs(e)\,kA5mA7~($\lambda$\,Boo) & \tablefootnote{Slight emission notch in H\,$\beta$. Mild $\lambda$\,Boo star - slight additional weakness in the \specline{Mg}{ii}{4481} line.} & 1 \\
445 & HD\,139787 & 9.77 & 3 & A4\,V &  &  \\
\bottomrule
\end{tabular}
}
\end{savenotes}
\end{table}

\setcounter{table}{0}
\renewcommand{\thetable}{A\arabic{table}}
\begin{table}
\centering
\caption{{\bf continued.} Spectral classes for the program stars.}
\begin{savenotes}
\resizebox{\columnwidth}{!}{%
\begin{tabular}{H l r c c l c}
\toprule
Run Num & Obj Name & \hspace{-2mm}V mag & \hspace{-2mm}Group\hspace{-2mm} & Class & \hspace{-1mm}Note & IR\\
\midrule
447 & HD\,140734 & 9.55 & 3 & A5\,IV & \tablefootnote{Great match to the A5\,IV standard, $\beta$\,Tri, except for the \specline{Ca}{i}{4226} line, which is much stronger than in the standard.} &  \\
1153 & HD\,141063 & 6.98 & 3 & kA2hA3mA5\,Va+ & \tablefootnote{Mild Am.} &  \\
1160 & HD\,141403 & 9.03 & 3 & A1\,Vbs &  & 1 \\
1154 & HD\,141442 & 8.74 & 3 & A1\,Va &  &  \\
1155 & HD\,141444 & 8.94 & 3 & A0\,Va & \tablefootnote{The \specline{Mg}{ii}{4481} line is weak, but this is not a $\lambda$\,Boo star, since the K line is normal.} &  \\
430 & HD\,141569 & 7.12 & 1 & A1\,Vn kB9mB9~$\lambda$\,Boo &  & 1 \\
1156 & HD\,141576 & 9.04 & 3 & A0.5\,Vas &  &  \\
1162 & HD\,141905 & 8.3 & 3 & A2\,Va+n &  &  \\
1157 & HD\,142404 & 9.14 & 3 & A1.5\,Vas & \tablefootnote{Slightly shallow H cores.} &  \\
446 & HD\,142524 & 9.59 & 3 & A1\,Van &  &  \\
452 & HD\,142666 & 8.82 & 1 & F0\,V~shell & \tablefootnote{Very deep H-line cores.} & 1 \\
1163 & HD\,142703 & 6.12 & 0 & F1\,Vs\,kA1.5mA1.5~$\lambda$\,Boo &  &  \\
1158 & HD\,142705 & 7.74 & 3 & A1\,Vann &  & 1 \\
460 & HD\,142931 & 9.79 & 3 & A1.5\,Vas &  &  \\
1167 & HD\,142994 & 7.17 & 0 & hF2\,V\,kA5mA5~($\lambda$\,Boo) &  & 1 \\
1168 & HD\,143511 & 8.31 & 3 & A1\,Vas & \tablefootnote{Great match to the standard star.} &  \\
495 & HD\,143567 & 7.19 & 3 & B9\,Van & \tablefootnote{Very slight weakness in the \specline{Mg}{ii}{4481} line, and the \ion{Ca}{ii} K line is a little weak, but B9 is too early to claim weak metal lines.} &  \\
494 & HD\,143600 & 7.33 & 3 & B9\,Vn & \tablefootnote{Slightly shallow H cores.} & 1 \\
493 & HD\,143715 & 7.14 & 3 & A1.5\,IVs & \tablefootnote{Weak H cores.} & 1 \\
1164 & HD\,143747 & 8.4 & 3 & A1\,IVn &  &  \\
1166 & HD\,143822 & 9.39 & 3 & A1\,Vbn &  &  \\
492 & HD\,143956 & 7.77 & 3 & B9\,Van &  & 1 \\
491 & HD\,144254 & 7.78 & 3 & A1\,Van~kA0.5 &  & 1 \\
490 & HD\,144273 & 7.54 & 3 & B9\,Vn &  & 1 \\
489 & HD\,144569 & 7.9 & 3 & A1\,Vas\,mA0.5 & \tablefootnote{Metal lines (except K line) are slightly weak. Not a $\lambda$\,Boo star.} & 1 \\
488 & HD\,144586 & 7.81 & 3 & A1\,IV-V\,kB9 & \tablefootnote{H lines well-matched at A1\,IV-V, not at A0\,IV-V. But trace He suggests A0. Could alternatively be a low-luminosity late-B star, e.g. B9.5\,Vbn. The \specline{Mg}{ii}{4481} line is weak, but rotation is very rapid, so probably not a $\lambda$\,Boo star.} & 1 \\
454 & HD\,144668 & 7.05 & 1 & A9\,V~shell & \tablefootnote{Very deep H-line cores.} & 1 \\
487 & HD\,144925 & 7.78 & 3 & A0\,Vn & \tablefootnote{Slightly shallow H cores.} & 1 \\
486 & HD\,144981 & 8.04 & 3 & A0.5\,Vn &  & 1 \\
485 & HD\,145188 & 7.06 & 3 & B9.5\,Vb &  & 1 \\
1169 & HD\,145631 & 7.6 & 3 & B9\,Vann &  & 1 \\
483 & HD\,146706 & 7.55 & 3 & B9\,Van &  & 1 \\
482 & HD\,147010 & 7.4 & 3 & B8:\,Vp~SrTi\,SiEu &  & 1 \\
484 & HD\,147046 & 7.8 & 3 & A2\,IVn & \tablefootnote{Excellent match, except for slightly shallow H cores.} & 1 \\
442 & HD\,148036 & 9.62 & 3 & F0.5\,Vn kA6mA6 & \tablefootnote{Not a $\lambda$\,Boo star.} &  \\
1170 & HD\,148534 & 9.02 & 3 & A2\,Vas &  &  \\
1171 & HD\,148563 & 8.72 & 3 & A2\,Va &  & 1 \\
1174 & HD\,148638 & 7.9 & 3 & A2\,IV-Vn &  & 1 \\
1175 & HD\,149130 & 8.48 & 3 & F1\,Vp Sr & \tablefootnote{Sr lines are strong. The \specline{Mg}{ii}{4481} is slightly weak. Not a $\lambda$\,Boo star.} & 1 \\
1173 & HD\,149151 & 8.12 & 3 & A0\,IV-V~SrSi &  &  \\
1172 & HD\,150035 & 8.71 & 3 & A2\,IVp~SrCrEu &  & 1 \\
455 & HD\,150193 & 8.79 & 1 & A3\,Va(e) & \tablefootnote{Slight emission notch in H\,$\beta$.} & 1 \\
467 & HD\,151873 & 9.1 & 2 & B9\,III~shell & \tablefootnote{Classical shell star.  Strong lines of \ion{Fe}{ii}, deep absorption cores in H lines, emission notch in H\,$\beta$.} & 1 \\
480 & HD\,153747 & 7.42 & 0 & A6\,Vn\,kA0mA0~$\lambda$\,Boo & \tablefootnote{An extreme $\lambda$\,Boo star.} &  \\
461 & HD\,153948 & 9.54 & 3 & B9\,IVp~SiSrCrEu &  & 1 \\
481 & HD\,154153 & 6.18 & 3 & F1.5\,Vs\,kA4mA4~$\lambda$\,Boo &  & 1 \\
1179 & HD\,154751 & 8.96 & 3 & A3\,IV &  &  \\
1180 & HD\,154951 & 8.78 & 3 & F2\,V\,kA6mA6~(($\lambda$\,Boo)) & \tablefootnote{A very marginal $\lambda$\,Boo star.} &  \\
1192 & HD\,155127 & 8.38 & 3 & kA0hA5mF0\,II &  &  \\
451 & HD\,155397 & 9.53 & 3 & F2\,Vs &  & 1 \\
1191 & HD\,156300 & 8.65 & 3 & A1:\,IIIp~EuCr(Sr) &  &  \\
1190 & HD\,156974 & 9.39 & 3 & A0.5\,IVs &  & 1 \\
1189 & HD\,157170 & 7.97 & 3 & kA0hA1mA2\,V & \tablefootnote{Possibly a very mild and early Am star.} &  \\
1188 & HD\,157184 & 9.48 & 3 & A1\,V &  &  \\
441 & HD\,157389 & 9.98 & 3 & B9\,IV-Vn &  & 1 \\
1187 & HD\,158681 & 8.22 & 3 & B6\,IV: & \tablefootnote{H cores are too deep for B5\,V, and wings are too deep for B5\,III. He lines are slightly weaker than B5, so B6\,IV is the best match.} &  \\
1181 & HD\,158830 & 8.97 & 3 & A1\,IV-V (shell) & \tablefootnote{\specline{Fe}{ii}{4232} is slightly enhanced, as well as the one line of the \ion{Fe}{ii}~(42) multiplet that is visible in this spectrum, suggesting a shell. H line cores are also quite deep, more so than can be explained by slow rotation.} & 1 \\
469 & HD\,159014 & 9.64 & 2 & B7\,IV-V(e) & \tablefootnote{Emission in core of H\,$\beta$, infilling in H\,$\gamma$. \ion{He}{i} slightly weak for B7, so may be B7.5.} & 1 \\
479 & HD\,160461 & 7.51 & 3 & A1.5\,IVn &  &  \\
1185 & HD\,161576 & 9.26 & 3 & A4\,Vn &  &  \\
1184 & HD\,161595 & 9.17 & 3 & A1\,Vas &  & 1 \\
478 & HD\,162220 & 6.66 & 3 & B9\,IVn &  & 1 \\
458 & HD\,163296 & 6.85 & 1 & A3\,Vae\,kA1mA1 & \tablefootnote{Not a $\lambda$\,Boo star, since the \specline{Mg}{ii}{4481} line is normal.} & 1 \\
\bottomrule
\end{tabular}
}
\end{savenotes}
\end{table}

\setcounter{table}{0}
\renewcommand{\thetable}{A\arabic{table}}
\begin{table}
\centering
\caption{{\bf continued.} Spectral classes for the program stars.}
\begin{savenotes}
\resizebox{\columnwidth}{!}{%
\begin{tabular}{H l r c c l c}
\toprule
Run Num & Obj Name & \hspace{-2mm}V mag & \hspace{-2mm}Group\hspace{-2mm} & Class & \hspace{-1mm}Note & IR\\
\midrule
465 & HD\,163921 & 9.52 & 3 & A9.5\,Vn & \tablefootnote{Intermediate between the A9\,Vn standard, 44\,Cet, and the F0\,V standard, HD23585, broadened to $v\sin i  = 150$\,km\,s$^{-1}$.} & 1 \\
475 & HD\,168740 & 6.12 & 0 & A9\,Vs\,kA2mA2~$\lambda$\,Boo & \tablefootnote{Classic $\lambda$\,Boo star.} & 1 \\
472 & HD\,168947 & 8.11 & 0 & A9\,Vs\,kA3mA4~($\lambda$\,Boo) & \tablefootnote{Cores are too narrow for an earlier giant (e.g. A4 III-IV). The \specline{Mg}{ii}{4481} line is weak.} &  \\
471 & HD\,169142 & 8.16 & 3 & F1\,Vs\,kA4mA5~($\lambda$\,Boo) & \tablefootnote{The \specline{Ca}{i}{4226} line is strong while \specline{Mg}{ii}{4481} is weak. A mild $\lambda$\,Boo star.} & 1 \\
473 & HD\,169346 & 9.27 & 3 & A8\,V\,kA6mA6~($\lambda$\,Boo) & \tablefootnote{The \specline{Mg}{ii}{4481} line is rather weak, but not much difference between h and km types. A mild $\lambda$\,Boo star. This target was observed twice and each spectrum was classified independently, arriving at similar classifications. The \specline{Sr}{ii}{4077} line is unusually strong in one of the spectra, but the \specline{Sr}{ii}{4215} line is normal.} &  \\
476 & HD\,171013 & 8.6 & 3 & F2\,Vs\,kA8mA7 & \tablefootnote{Metal weak, not clearly $\lambda$\,Boo in nature.} & 1 \\
477 & HD\,176386 & 7.21 & 3 & B9\,Vbs & \tablefootnote{Could serve as spectral standard for B9\,Vbs.} & 1 \\
1178 & HD\,176387 & 8.94 & 4 & A5\,II-III\,kA0mA0 & \tablefootnote{Blue horizontal branch star? \specline{Mg}{ii}{4481} is weak, although not with respect to A5\,II line ratios.} & 1 \\
1182 & HD\,184779 & 8.9 & 0 & F0\,V\,kA5mA6~($\lambda$\,Boo) &  &  \\
1193 & HD\,188230 & 8.18 & 3 & A0\,Vbn &  &  \\
1014 & HD\,261520 & 10.11 & 1 & B5\,II-IIIe~shell & \tablefootnote{Emission in H\,$\beta$ (in an inverted `w' shape), strong absorption in Si~4128-30 and the \ion{Ca}{ii} K line.} & 1 \\
1015 & HD\,288947A & 11.13 & 2 & B7\,IV-Ve & \tablefootnote{Emission partially fills H\,$\beta$.} & 1 \\
1025 & HD\,290469 & 9.87 & 3 & A4\,Vs &  & 1 \\
1024 & HD\,290470 & 9.77 & 3 & A2.5\,V &  & 1 \\
1028 & HD\,290516 & 9.51 & 3 & B9\,Vbn &  &  \\
1034 & HD\,290666 & 10.03 & 3 & A0\,Van &  & 1 \\
1029 & HD\,290684 & 9.7 & 3 & A2\,V\,kA1 &  &  \\
1016 & HD\,292895 & 11.16 & 2 & B8\,Vn & \tablefootnote{Slightly noisy spectrum.} & 1 \\
1027 & HD\,294054 & 9.6 & 3 & kA0hA0mA1\,Vbn & \tablefootnote{Possibly a mild Am star, with the K line weaker than the metals, but also definitely a rapid rotator.} &  \\
1022 & HD\,294103 & 9.7 & 3 & A1\,Van &  & 1 \\
1035 & HD\,294253 & 9.67 & 3 & A0\,Va\,kB8.5mB9~($\lambda$\,Boo) &  & 1 \\
1017 & HD\,296192 & 10.21 & 2 & B7.5\,IV-Ve & \tablefootnote{Emission partially fills H\,$\beta$.} & 1 \\
1061 & HD\,304838 & 9.87 & 3 & B8.5\,V &  &  \\
1070 & HD\,307860 & 8.28 & 3 & B8.5\,Vnn &  & 1 \\
426 & HD\,308889 & 10.64 & 2 & B6\,V(e) &  & 1 \\
427 & HD\,309344 & 10.85 & 2 & A4\,III-IV &  & 1 \\
466 & HD\,314915 & 11.31 & 2 & A9\,Vn\,kA5mA5~($\lambda$\,Boo) & \tablefootnote{Very rapid rotation. Even so, the \specline{Mg}{ii}{4481} line is weak.} & 1 \\
1186 & HD\,318093 & 9.71 & 3 & A0\,Va+s &  & 1 \\
464 & HD\,318099 & 9.86 & 3 & A0\,Van & \tablefootnote{Slightly shallow H cores.} &  \\
463 & HD\,318127 & 9.82 & 3 & A1\,IVn &  & 1 \\
468 & HD\,320460 & 10.63 & 2 & B5\,IIIe & \tablefootnote{Emission core in H\,$\beta$, infilling in H\,$\gamma$, and possibly H\,$\delta$.} & 1 \\
1183 & HD\,320765 & 8.73 & 3 & A1\,V &  & 1 \\
462 & HD\,322663 & 9.77 & 3 & A1\,IVn &  & 1 \\
1299 & BD+00\,2757 & 10.64 & K2 & F5\,V:\,mF2gF5 &  &  \\
367 & BD-02\,2182 & 9.71 & 2 & B2\,Vn &  & 1 \\
1040 & BD-08\,1151 & 9.82 & 3 & A2\,IV-V &  & 1 \\
1041 & BD-11\,1239 & 9.7 & 3 & A7\,V\,kA3mA3~(($\lambda$\,Boo)) &  & 1 \\
374 & BD-11\,1762 & 10.05 & 3 & B2\,IV-Vn &  & 1 \\
366 & BD-15\,1548 & 9.88 & 1 & B7\,IIIe~He-wk &  & 1 \\
1177 & BD-15\,4515 & 9.97 & 4 & F2\,V\,kA4mA6~$\lambda$\,Boo &  & 1 \\
373 & CD-31\,4428 & 9.86 & 3 & A2\,Van &  &  \\
387 & CD-37\,3833 & 9.92 & 3 & A2\,Vn\,kA0 &  & 1 \\
372 & CD-48\,3541 & 9.67 & 3 & A2\,Vn\,kA0mA1 & \tablefootnote{Not a $\lambda$\,Boo star, just slightly metal weak} & 1 \\
385 & CD-55\,2595 & 10.31 & 2 & B1\,Ve & \tablefootnote{Spectrum contaminated, H\,$\beta$ partially filled with emission.} & 1 \\
393 & CD-58\,3782 & 9.79 & 3 & A3\,Van &  &  \\
1055 & CD-59\,1764 & 9.65 & 3 & A0.5\,V &  & 1 \\
1043 & CD-60\,1932 & 9.93 & 3 & A0\,Vnn & \tablefootnote{Rotating more rapidly than the high $v\sin i$  A0\,V standard.} & 1 \\
1056 & CD-60\,1956 & 9.65 & 3 & A0.5\,V &  &  \\
1057 & CD-60\,1986 & 9.57 & 3 & A2\,Van &  &  \\
1105 & CD-60\,4157 & 9.38 & 3 & A1\,Van &  & 1 \\
440 & CD-60\,6017 & 9.63 & 3 & A8\,IV-V &  &  \\
412 & CD-60\,6021 & 9.73 & 3 & B7\,IVn &  &  \\
375 & CPD-20\,1613 & 10.0 & 3 & A0.5\,V\,kB9.5 &  &  \\
406 & CPD-58\,3071 & 9.85 & 3 & A3\,Va &  &  \\
394 & CPD-58\,3106 & 9.76 & 3 & A1.5\,Vn &  &  \\
395 & CPD-58\,3138 & 9.73 & 3 & A1.5\,Vs &  & 1 \\
1125 & IK\,Hya & 10.23 & 4 & B7\,II\,kA3mA3 & \tablefootnote{Known RRLyr variable.} &  \\
456 & KK\,Oph & 10.99 & 1 & A9:e & \tablefootnote{Emission in H\,$\beta$, \ion{Ca}{ii}\,K\,and\,H, and some other metallic lines. Probably pre main sequence, or possibly an RS\,CVn variable. This is KK\,Oph, a known Herbig Ae/Be star.} & 1 \\
457 & NGC\,6383\,22 & 12.49 & 1 & F1\,V(e)\,kA6mA6~$\lambda$\,Boo? & \tablefootnote{The \ion{Ca}{ii}\,K line has a peculiar profile (broad but shallow). Enhanced G-band absorption suggests this is a composite spectrum. The \specline{Mg}{ii}{4481} line is slightly weak. A high-resolution spectrum is needed to determine if this is a mild $\lambda$\,Boo star or a composite instead.} & 1 \\
363 & T\,Ori & 11.25 & 1 & A8\,Vne kA1mA2~$\lambda$\,Boo & \tablefootnote{H\,$\beta$ in emission, emission core in H$\gamma$.} & 1 \\
1012 & V1012\,Ori &  & 1 & F3\,V\,kA5n~composite & \tablefootnote{Might be a triple system: both line veiling, indicating a hotter companion, and a strong G-band red edge, indicating a cooler companion, are evident.}\\
444 & V748\,Cen & 11.93 & 2 & A0e & \tablefootnote{Most lines are in emission, H\,$\beta$ strongly so. SpT from Ca II K line.} & 1 \\
376 & V\,Lep & 9.71 & 3 & F0.5\,V(n)\,kA8mA9 & \tablefootnote{Not a $\lambda$\,Boo star, just slightly metal weak.} & 1 \\
\bottomrule
\end{tabular}
}
\end{savenotes}
\end{table}

\end{document}